\documentclass[11pt,twoside]{article}
\usepackage{asp2004}
\usepackage{psfig}
\usepackage{epsf}
\usepackage{graphics}
\usepackage{lscape}
\def\edcomment#1{\iffalse\marginpar{\raggedright\sl#1\/}\else\relax\fi}
\reversemarginpar

\begin{document}
\title{Faint X-ray sources in globular clusters in the XMM-Newton and Chandra era}
\author{Natalie A. Webb}
\author{Didier Barret}
\affil{Centre d'Etude Spatiale des Rayonnements, 9 av. du Colonel Roche, 31028 Toulouse, France}

\begin{abstract}

X-ray imaging and spectroscopy can be used to probe the binary content of
globular clusters. Binaries are thought to play a key role in the
dynamical evolution of the clusters by serving as an internal source of
energy which counters the tendency of their cores to collapse. Various
kinds of binaries are found in clusters in X-rays, including BY Dra or RS
CVn systems, Cataclysmic Variables and low mass X-ray binaries as well as
millisecond pulsars. In this review, we will present the most recent X-ray
observations of globular clusters as performed with XMM-Newton and
Chandra. We will illustrate the power of X-ray observations to separate the
different X-ray source populations, and emphasize how the results are
important in understanding the binary formation processes.

\end{abstract}

\section{Introduction}

It is thought that through dynamical evolution, globular cluster cores
become increasingly more concentrated \citep{heno61} and eventually
collapse.  This core collapse should occur within 
$\sim$10$^8$-10$^{10}$ years unless additional energy is injected into the
central regions of the cluster.  This would indicate that
many Galactic globular clusters should have already undergone core
collapse.  However, we observe that many have not yet done so, see
e.g. \cite{harr96}.

Both primordial binary systems and those formed due to encounters
should exist in globular clusters due to the dense environments.
Binaries can subsequently encounter cluster stars and in doing so, if
the binding energy of the binary is larger than the kinetic energy of
the cluster star, harden (become tighter) \citep{hegg75}.  The third
body leaves the encounter with more energy than before the interaction
\citep[see e.g.][for reviews]{hut92,elso87}.  It is therefore
clear from this, that a population of binaries could be important in
helping to delay globular cluster core collapse. Binaries and their
descendants are thus valuable objects with which to complete
the study of stellar and globular cluster evolution, stellar dynamics,
binary formation and evolution, as well as the study of compact
objects and the binaries themselves, see e.g. \cite{verb04,hut92} for
extensive reviews of these topics.

X-ray sources were first detected in Galactic globular clusters in the
early 1970's using the {\em Uhuru} and {\em OSO-7} observatories
\citep{giac72,giac74,clar75}.  Following the launch of the
{\em EINSTEIN} observatory in 1978, \cite{hert83} surveyed 71 Galactic
globular clusters and showed that there were in fact two distinct
populations of globular cluster X-ray sources.  The bright sources, with L$_x\ >$ 10$^{36}$ erg s$^{-1}$, were proposed to be
accreting neutron star systems.  The nature of the faint sources, with
L$_x\ _\sim ^<$ 10$^{34.5}$ erg s$^{-1}$, was unclear but it was
thought that they could also be binary systems containing a compact
object, but in this case a white dwarf.  \cite{verb84} then suggested
that the brighter of the faint sources, which they stated were too
bright to be cataclysmic variables (CVs), by comparing their
luminosities with Galactic plane CVs, could be quiescent soft X-ray
transients. \cite{marg87} also supported a hypothesis made by
\cite{hert83b}, that some of the faint X-ray sources could be
fore- and background superpositions of unrelated sources.  The
discovery of radio millisecond pulsars (MSPs), which are believed to be
descendants of neutron star X-ray binaries, in globular clusters
\citep{lyne87} was followed by the identification of the X-ray
counterparts \citep[e.g.][]{dann97}, indicating that some of the faint
X-ray sources could be MSPs. Bailyn, Grindlay, \& Garcia (1990) also
suggested that some of the faint sources could be active binaries,
such as RS~CVn binaries.

Observations with the {\em Rosat} observatory
\citep[e.g.][]{john94,john96b,verb01} dramatically increased the
number of known faint X-ray sources, although the number of
identifications of these sources remained limited.  The new generation
of X-ray telescopes e.g. {\it XMM-Newton} and {\it Chandra}, whose
sensitivity and angular resolution (respectively) far surpasses that
of previous X-ray observatories, has not only allowed us to detect but
also identify a large number of the faint X-ray source population.

Stellar encounters in globular clusters result in an equipartition of
energy.  Through virialization, the more massive stars,
such as degenerate objects and binary systems, are concentrated
towards the centre of the cluster.  The angular resolution of the {\it
Chandra} instruments can be as good as 0.4''
\citep{char00}, which makes it the ideal instrument to peer into the
centres of globular clusters that are either nearing or have undergone
core collapse.  The large majority of X-ray sources can thus be
resolved and unprecedented positional precision can obtained,
essential for multi-wavelength follow-up observations, which are in
turn crucial in determining the nature of the X-ray sources.  The {\it
XMM-Newton} observatory, thanks to its 58 grazing-incidence mirrors in
each of its three telescopes, is an extremely sensitive instrument
\citep{jans01}, ideal for observing faint X-ray sources.  Its angular
resolution of 6'' (Full Width Half-Maximum of its Point Spread
Function), limits its usefulness to less concentrated globular
clusters.  However, the large collecting area means that for even
faint sources ($\sim 1 \times 10^{-14}$ erg cm$^{-2}$ s$^{-1}$) in
less concentrated clusters, a useful X-ray spectrum can be obtained in
a short observation i.e. 15 kiloseconds and thus the nature of
certain sources derived from X-ray spectroscopy alone.

In this review we concentrate on {\it XMM-Newton}'s contribution in
the last five years, to understanding faint X-ray sources (and thus
binary systems) in Galactic globular clusters (see
Table~\ref{tab:xmmgcs} for clusters observed).  For articles oriented
towards {\it Chandra}'s contribution to this subject, we recommend,
amongst others, \cite{verb04,hein03,pool03,pool02,grin01}.  For
discourses on black hole(s) (binaries) and neutron star retention in
globular clusters, which are not discussed here, see
\cite{kalo04,ivan04,pods04,pfah02}.

\begin{table}
     \label{tab:xmmgcs} \begin{tabular}{cccccc} \hline
     \noalign{\smallskip} Name & alias & core & half-mass & Sources in
     core & Sources in half- \\ & & radius & radius & (Expected) &
     mass (Expected)\\ \hline NGC 3201 & & 87 & 161 & 2 (0.7$\pm$0.1)
     & 4 (1.8$\pm$0.4)\\ NGC~5139 & $\omega$~Cen & 155 & 258 & 11 (4)
     & 27 (9) \\ NGC~6205 & M~13 & 55 & 98 & 2 (0) & 5 (1) \\ NGC~6366
     & & 110 & 154 & 1 (1) & 3 (2) \\ NGC~6656 & M~22 & 85 & 198 & 3
     (1.0$\pm$0.1) & 6 (3.3$\pm$0.8) \\ NGC~6809 & M~55 & 170 & 173 &
     4 (3.5$\pm$1)& 5 (3.5$\pm$1)\\
\hline
  \end{tabular}
  \caption[]{Globular clusters observed by {\it XMM-Newton}.  Physical characteristics (core and half mass-radii (in arcseconds)) and the number of sources detected, along with the expected number of sources if no cluster was present (i.e. fore- and background sources) for the core and half-mass radii.}
 \end{table}

\section{Binary formation}

Globular clusters are extremely dense environments, with as many as
10$^5$ stars pc$^{-3}$ \citep{davi02},
thus stellar encounters, which are extremely rare in lower density
regions, can occur in globular clusters on time-scales comparable with
or less than the age of the Universe.  This would indicate that many
Galactic globular cluster stars have undergone at least one encounter
in its lifetime.  Encounters between stars is one way in which
binaries can be produced.  The encounter rate ($\Gamma$) due to tidal capture
\citep{fabi75} is proportional to the encounter cross-section ($A$), the
relative velocity of the stars ($v$) and the number density of stars
in the cluster (core).  For encounters between a compact object and an
ordinary cluster star, we require $n_c$ compact objects and $n_s$
cluster stars per unit volume, see \cite{hut83,verb03} and
Eq.~\ref{eq:encounter}.

\begin{equation}
\label{eq:encounter}
\Gamma \propto \int n_c n_s A v dV \propto \int \frac{n_c n_s R}{v}  dV \propto \frac{\rho^2 r_c^3}{v}  R
\end{equation}

where $\rho$ is the central mass density and $r_c$ the core radius.

An encounter between a binary and either a single star or a
binary system would more readily occur as the cross sections
are significantly larger, thus increasing the likelihood of an
encounter.  Wide (soft) binaries would be broken up in such an
interaction, but tight (hard) binaries interacting with single stars
could result in triple systems, a merged binary, a common envelope
system or more likely a fly-by or an exchange binary.  Such encounters
were thought to explain the over-abundance of binaries, in particular
neutron star binaries in globular clusters, compared to the number in
the field \citep{clar75b}. 

A great deal of work has been done on binary formation in globular
clusters and we refer the reader to several papers and reviews not
already referenced in this section for further details
e.g. \cite{davi97,dist94,verb88,hill75}.  However, for the formation of most
types of binary, it is only in recent years, with the launch of the
new generation of X-ray observatories, that we are beginning to be able
to observationally test the theory (and simulations) that have been
developing over the last 30 years.  These observations are discussed
in Sections~\ref{sec:formqnslmxbs} and \ref{sec:formcvs}

\section{Neutron star low mass X-ray binaries}
\label{sec:nslmxbs}

Bright X-ray sources in globular clusters were shown to be neutron
star low mass X-ray binaries (NSLMXBs) following X-ray observations
with the early X-ray satellites and studying their X-ray bursts
\citep{cani75,hert83}.  \cite{verb87} showed that there was a
correlation between the number of NSLMXBs and the globular cluster
encounter rate.  This indicated that most NSLMXBs were formed through
encounters.

Amongst the ever growing population of faint X-ray sources in globular
clusters, a subset were found to have very soft X-ray spectra
that are well fitted by a neutron star hydrogen atmosphere model
\citep[e.g.][]{pavl92,zavl96}.  The model fits indicate radii that are
consistent with those expected from either a neutron star
\citep[approximately 10 km e.g.][]{latt01} or a polar cap being heated
by material accreted from a companion star and channelled towards the
pole via a magnetic field \citep{hein03}.  These sources have X-ray
luminosities between $\sim$10$^{32}$ and a few 10$^{33}$ erg s$^{-1}$,
which are consistent with X-ray luminosities of quiescent NSLMXBs
(qNSLMXBs) in the disc \citep{nara02}.  This supports the idea that
these sources are also qNSLMXBs.  Indeed \cite{hagg04} recently found
the probable optical counterpart to the X-ray source of the proposed
qNSLMXB in $\omega$ Centauri.  The optical source, found with the
Advanced Camera for Surveys (ACS) on the Hubble Space Telescope is a
blue source with H$_\alpha$ emission, implying that there is some
accretion resulting in an accretion disc, although the faintness of
the disc suggests a very low rate of accretion.  They constrain the
mass of the counterpart to be M$<$0.14M$_\odot$, if it is a
main-sequence star.  All of these points are consistent with the idea
that this is the optical counterpart to a qNSLMXBs.

\begin{figure}
\plotfiddle{webb2_f1.ps}{0cm}{270}{24}{24}{-180}{20}
\plotfiddle{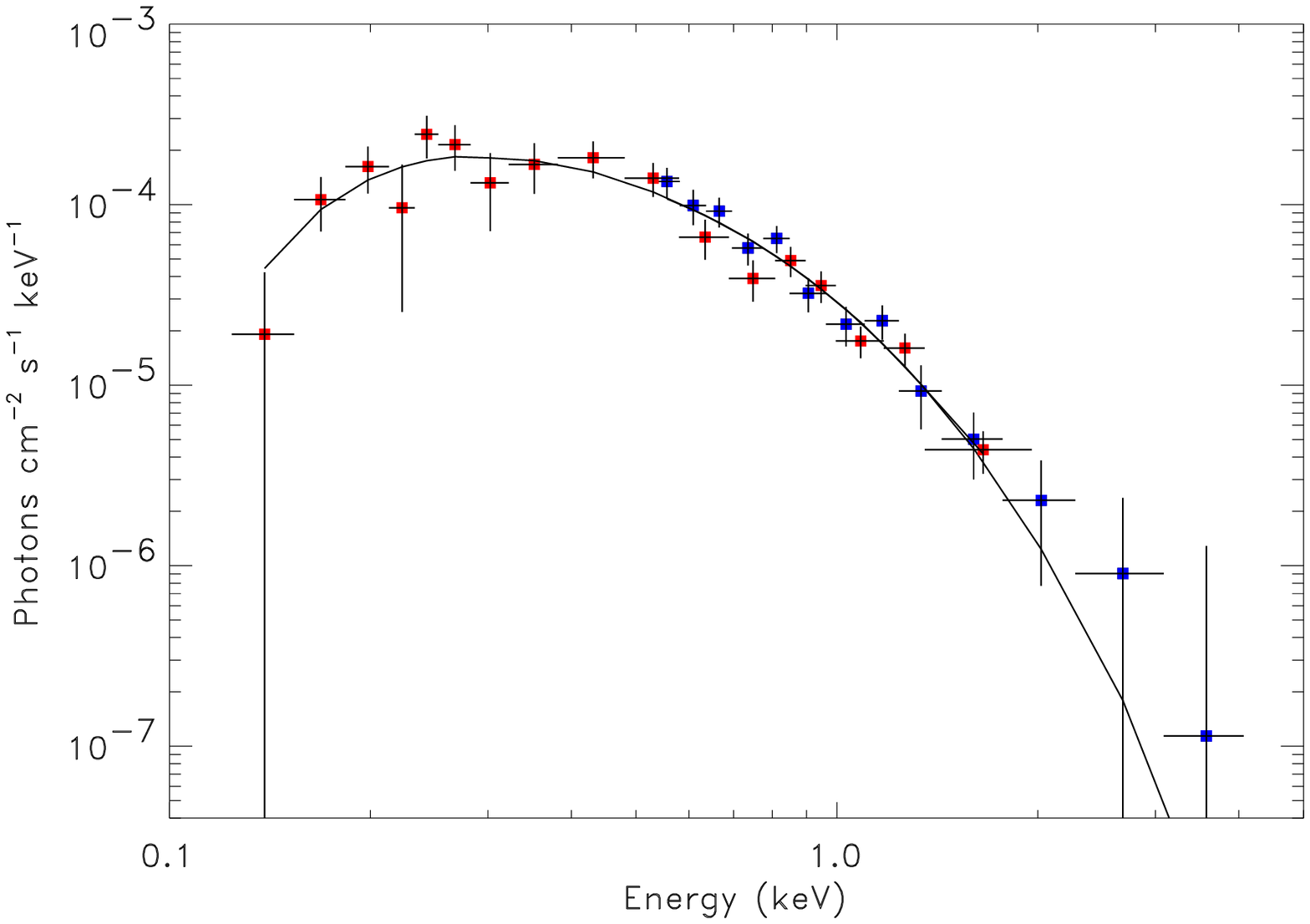}{0cm}{0}{33}{33}{-10}{-130}

\vspace*{3.5cm}
      \caption{Left: The unfolded EPIC spectra of the
      qNSLMXB candidate in $\omega$ Centauri
      \citep{gend03a}. The solid line indicates the hydrogen
      atmosphere model fit. Right: The
      unfolded EPIC-PN (diamonds) and ROSAT-PSPC (squares) spectra of
      the qNSLMXB candidate in M~13 \citep{gend03b} and the
      best fit hydrogen atmosphere model. }

       \label{fig:qNSLMXBspec}
   \end{figure}

So far 28 X-ray sources with spectra indicative of qNSLMXBs
have been found using either {\it XMM-Newton} or {\it Chandra}, see
\cite{hein03} for a list of the best fitting parameters to the X-ray
spectra, luminosities etc. The X-ray spectra for the proposed
qNSLMXBs detected with {\it XMM-Newton} in $\omega$ Centauri
(Gendre, Barret, \& Webb 2003a) and M~13 ({\em Rosat} data also
included in plot, Gendre, Barret, \& Webb 2003b) can be seen in
Figure~\ref{fig:qNSLMXBspec}.

\subsection{Neutron star X-ray binary formation in globular clusters}
\label{sec:formqnslmxbs}

As the faintest detection limit of all the {\it XMM-Newton} and {\it
Chandra} observations of globular clusters ($\sim$10$^{29}$-10$^{31}$
erg s$^{-1}$) is considerably fainter than the minimum luminosities of
qNSLMXBs ($\sim$10$^{32}$ erg s$^{-1}$), we can consider that
the populations of qNSLMXBs detected in each globular cluster
are (almost) complete \citep[although source confusion could mar the
completeness][or some qNSLMXBs may show hard X-ray spectra
and not be accounted for in this analysis, see
Section~\ref{sec:xrayspec}]{hein03}.

\begin{figure}
\plotfiddle{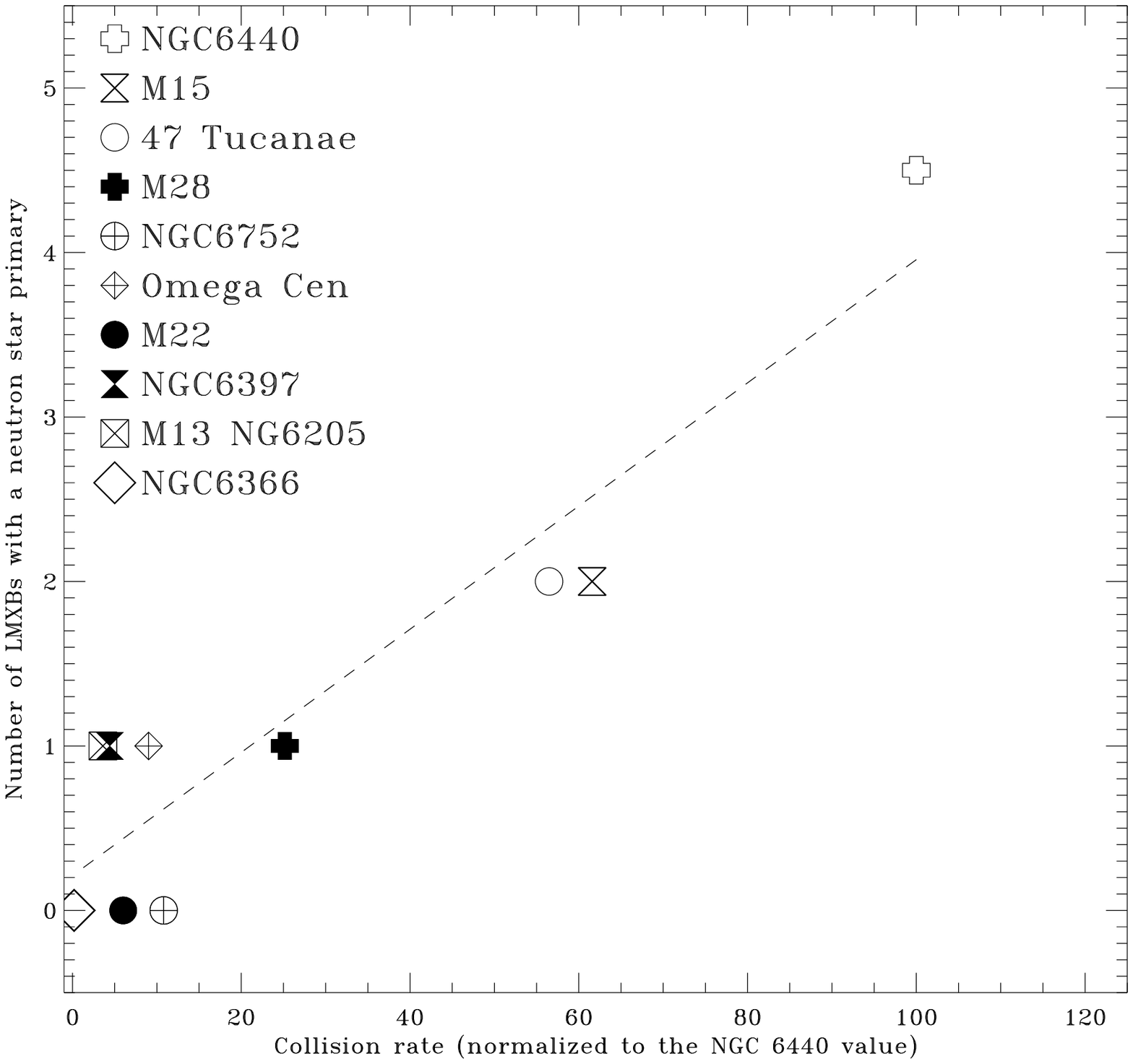}{0cm}{0}{35}{33}{-200}{-185}
\plotfiddle{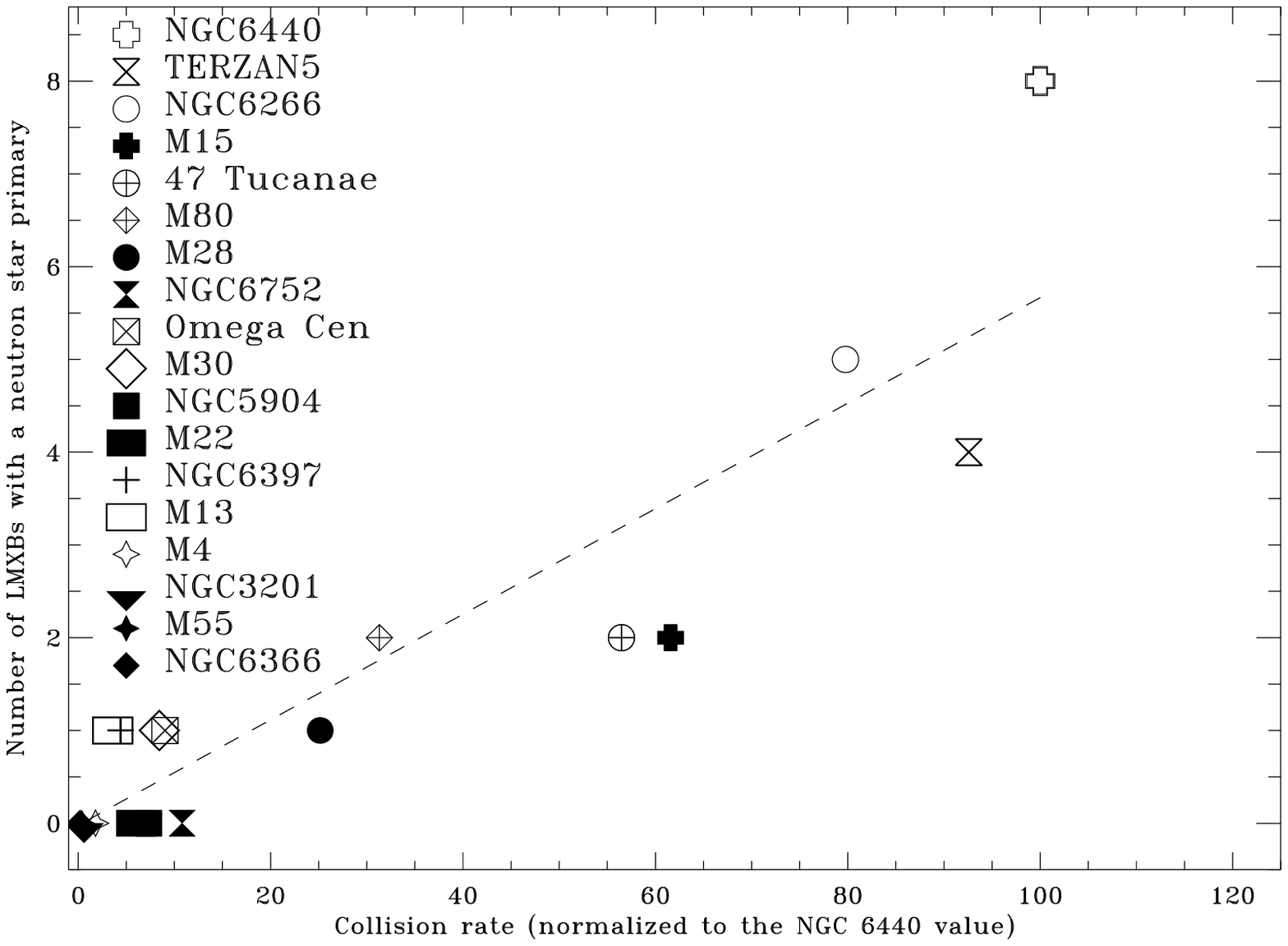}{0cm}{0}{40}{48}{-30}{-305}

\vspace*{4.5cm}
      \caption{Left: The number of qNSLMXBs versus collision
      rate ($\rho_0^{1.5} r_c^2$), shown with a linear fit, $n_{\rm
      qNSLMXB} \sim 0.04 \times \rho_0^{1.5} r_c^2 + 0.2$. The
      collision rates have been normalized to 100 for NGC~6440
      \citep{gend03b}.  Right: The number of qNSLMXBs versus
      collision rate for all 18 globular clusters that have been
      observed with {\it XMM-Newton} or {\it Chandra} by mid 2004. }

       \label{fig:qNSLMXBrel}
   \end{figure}

Plotting the number of qNSLMXBs found with soft spectra
against the globular cluster collision rate, \cite{gend03b} found that
a correlation existed for the 10 globular clusters (containing 10-11
qNSLMXBs) that had been observed by either {\it XMM-Newton}
or {\it Chandra} by 2003, see Fig.~\ref{fig:qNSLMXBrel} left hand
side.  The data could be fitted by a linear fit, although the numbers
involved are low and therefore the statistical significance is low.
However, the observations do support the idea that qNSLMXBs
are formed through encounters, in a similar way to the NSLMXBs.
\cite{pool03} subsequently analysed 14 globular clusters, containing a
total of 19-22 qNSLMXBs and found that this analysis also supported
the idea that qNSLMXBs are formed in large part through encounters.
They found that a power law fit with an index of 0.97$\pm$0.5 fitted
the number of LMXBs versus the collision rate, consistent with a
linear relationship. \cite{hein03} also support this hypothesis, using
the density weighting method of \cite{john96a} and using an estimate
for the incompleteness of qNSLMXBs detected in certain clusters.  In
Fig.~\ref{fig:qNSLMXBrel}, right hand side, we plot the 18 globular
clusters that have been observed by either {\it XMM-Newton} or {\it
Chandra} by mid 2004, which include approximately 28 qNSLMXBs see
\cite{hein03} and references therein, \cite{webb04b} and \cite{bass04}
for the 18 globular clusters and their qNSLMXBs.  The numbers involved
are still fairly small, but the fit is still consistent with a linear
relationship.

\subsection{The population of globular cluster neutron star X-ray binaries}

Using results presented above, \cite{pool03} estimate that there are
approximately 100 LMXBs residing in Galactic globular clusters. Of
these 100 LMXBs, there are 13 persistently or transiently
bright LMXBs, many of which have been known for almost 20 years. With
the increase in the number of detected qNSLMXBs over the last 5 years,
thanks to {\it XMM-Newton} and {\it Chandra}, we have already
recovered about 40\% of the population.  The entire population
will be very useful for testing models of cluster evolution
and LMXB formation.

\subsection{The neutron star equation of state}

This population is also useful for other observational tests.  As
stated above, qNSLMXBs have spectra that are, in general,
well fitted by neutron star hydrogen atmosphere models.  These models
have only 5 parameters.  Two of the parameters are fairly well
constrained, as the binary is located in a globular cluster: the
distance (known to a few percent thanks to {\it Hipparcos}) and the
interstellar absorption.  The three remaining parameters are the mass
of the neutron star and its temperature and radius.  Assuming a
canonical mass for the neutron star, we can constrain the other two
parameters.  For the two qNSLMXBs detected by {\it
XMM-Newton} we find $T^{\scriptscriptstyle
\infty}_{\scriptscriptstyle eff}$ = 67 $\pm$ 2 eV and $R_{\infty}$ =
13.6 $\pm$ 0.3 km. The unabsorbed luminosity is $L$ = 3.2 $\pm$ 0.2
$\times 10^{32}$ erg s$^{-1}$ (0.1-5.0 keV), for the source in
$\omega$ Cen.  We find T$_{\infty}=76\pm3$ eV and
R$_{\infty}=12.8\pm0.4$ km and an X-ray luminosity of
$7.3\pm0.6\times10^{32}$ erg s$^{-1}$ (0.1-5.0 keV) for the source in
M~13, assuming a mass of 1.4M$_\odot$ in both cases.  If the X-ray
emission observed is indeed due uniquely to the neutron star cooling,
then the measurements for the temperature and
radius using {\it XMM-Newton} are very accurate.

By detecting the optical counterpart, which is now starting to be
possible thanks to the small errors in the positions of the X-ray
sources (in particular the {\it Chandra} sources) and good angular
resolution obtained with optical telescopes such as the {\em Hubble
Space Telescope} ({\em HST}) and the {\em Very Large Telescope} ({\em
VLT}) we can eventually derive the mass function of the binary, using
optical spectroscopy and Kepler's laws.  Using optical spectroscopy to
determine the spectral type and thus the mass of the companion, we can
deduce the mass of the compact object and thus ultimately constrain
both the mass and radius of the neutron star.  If the
mass and radius can be determined for a substantial part of the
proposed population of NSLMXBs in Galactic globular clusters, this
will place very strict constraints on the neutron star equation of
state, which is important to our understanding of the physics of
matter at ultra-high densities as well as the astronomical
implications, understanding: the core collapse of massive stars, the
supernova phenomenon and the existence and properties of neutron stars
\citep{vank04,latt01}.

\subsection{Understanding the nature of neutron star X-ray binaries}
\label{sec:xrayspec}

The well constrained temperatures obtained using these models can be
used to estimate the cooling time. The exact cooling time depends on
the thermal conductivity of the crust, the core cooling processes and
the accretion history of the source \citep{wijn04}, but by assuming
the mass transfer rate during outburst, the time between outbursts can
be determined \citep[e.g.][]{rutl02}.  Observing a system going into outburst,
then fading into quiescence could help give us a handle on accretion
processes.

Several NSLMXBs, including two field qNSLMXBs: Cen X-4
\citep{asai96} and Aql X-1 \citep{camp98b} show  hard spectral 
components that dominate the spectrum above 2 keV.  The nature of this
hard power-law tail is not yet understood.  It has been explained by
accretion onto the neutron star surface, accretion down to the
magnetospheric radius or non-thermal processes powered by the
rotational energy loss of a rapidly spinning neutron star \citep[][and
references therein]{camp98a}. Using the {\it XMM-Newton} 
qNSLMXBs spectra above 2-3 keV, we have been able to place a limit on
the contribution from a hard power law tail.  For the $\omega$ Cen
source, assuming a power law with photon index of 2, an upper limit of
10\% of the total flux (90\% confidence limit) was derived
\citep{gend03a}.  For the source in M~13, \cite{gend03b} found no 
evidence for a hard power-law contribution.  \cite{hein03} list 21
probable globular cluster qNSLMXBs and find that only one
(NGC 6440 CX1) requires a power law component in fitting the X-ray
spectrum and is the only qNSLMXBs in their sample to have
experienced a recorded outburst.  They suggest that the strength of
the power-law component may be a measure of continuing low-level
accretion. Indeed the qNSLMXB in Terzan 5 (EXO 1745-248),
which shows a hard power law spectrum with little evidence for a soft
component \citep{wijn04b} had previously been in a high state,
indicative of recent accretion.  Further observations of these
Galactic globular cluster qNSLMXBs should help in resolving
the nature and origin of the hard power law tail.

\section{Cataclysmic variables}
\label{sec:cvs}

A large number of cataclysmic variables are now being discovered in
Galactic globular clusters.  Of the 57 faint X-ray sources known in
Galactic globular clusters following the {\em Rosat} period
\citep{verb01}, only a few had been identified as CVs 
\citep[e.g.][]{grin95,cool98,cars00}.  With the hundreds of
new X-ray sources being detected in Galactic globular clusters
with the new generation of X-ray observatories, many new CVs are being
detected and identified.  Part of this is due to the excellent
sensitivity of {\it XMM-Newton}, as we can now obtain X-ray spectra
and lightcurves of these faint sources which show CV characteristics,
i.e. spectra well fitted by high temperature bremsstrahlung models and
lightcurves showing variability on different timescales (seconds to
days) \citep{rich96,erac91,osbo87}.

\begin{figure}
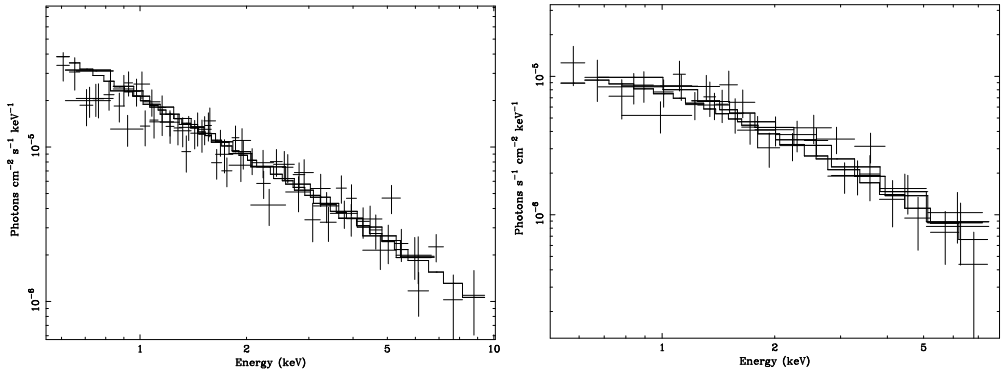

\plotfiddle{webb2_f5.ps}{0cm}{270}{28}{28}{-200}{10}
\plotfiddle{webb2_f6.ps}{0cm}{270}{28}{28}{-10}{35}

\vspace*{4.5cm}
      \caption{The EPIC-PN unfolded spectra of the two CV candidates located in the core of the globular cluster $\omega$ Centauri \citep{gend03a}. These spectra are shown with a best thermal bremsstrahlung fit. }

       \label{fig:CVspec}
   \end{figure}

Two of the core sources in $\omega$ Centauri, first detected by
\cite{cool95} were proposed to be CVs from the characteristics of
their optical counterparts, observed with the HST Wide Field Planetary
Camera 2.  The counterparts were found to have with strong H$_\alpha$
emission and a blue excess.  The fits to the X-ray spectra observed
with {\it XMM-Newton} support the CV hypothesis \citep{gend03a}.  The
spectra can be seen in Fig.~\ref{fig:CVspec} and the fits are
bremsstrahlung models with KT=23.0$^{\scriptscriptstyle
+39.0}_{\scriptscriptstyle -10.1}$ keV, $\chi^{\scriptscriptstyle
2}_{\scriptscriptstyle \nu}$=0.72 (31 degrees of freedom (dof)),
L$_{0.5-10.0 keV}$=6.7($\pm$0.3) $\times$10$^{32}$ erg s$^{-1}$ and
KT=18.7$^{\scriptscriptstyle +65.2}_{\scriptscriptstyle -10.0}$ keV,
$\chi^{\scriptscriptstyle 2}_{\scriptscriptstyle \nu}$=0.78 (17 dof),
L$_{0.5-10.0 keV}$=2.7($\pm$0.2) $\times$10$^{32}$ erg s$^{-1}$.
Analysing the X-ray colours, there is evidence for a small population
of CVs in this cluster.

Spectral fitting of a source found in the core of the globular cluster
M~22 using {\it XMM-Newton} data, coupled with optical colours,
indicates that this source is a CV \citep[KT=19.98$\pm$13.35 keV,
$\chi^{\scriptscriptstyle 2}_{\scriptscriptstyle \nu}$=1.57 (25 dof)
and L$_{0.2-10.0 keV}$ = 6.0$\times$10$^{31}$ erg
s$^{-1}$][]{webb04a}.  The source is also found to be variable on time
scales of hours. The two possible optical counterparts have U-band
magnitudes of 18.9 and 19.1 and U-V values of 0.4 and 0.3 respectively.  If
either of these is the counterpart, the L$_x$/L$_{opt}$ value
of $\sim$0.6 indicates a CV nature
\cite[e.g.][]{schw02,verb97}. There are a further 6-7 sources in this
cluster that also have X-ray and optical colours consistent with a CV
nature.  Follow-up spectroscopy of these sources taken with the VLT
UT3 (VIMOS) is currently being analysed and will help clarify the
uncertain nature of these sources.

X-ray spectral and temporal observations with {\it XMM-Newton} of the
globular clusters M~55 and NGC~3201, reveal one and three sources that
appear to be CVs.  X-ray and optical colours of a
second source in M~55 support the idea that there is a second CV in
this cluster \citep{webb04b}.
  
Many CVs have been detected using follow-up optical observations of
globular clusters observed with {\it Chandra}.  X-ray sources detected
with this satellite have positional errors of the order of 1''.  With
HST photometry of these sources, the optical counterparts can be
located.  Clusters which have revealed substantial populations of CVs
using this method are: NGC~6397 in which 6-7 sources have been
optically identified as CVs
\citep{grin01b}, six CVs have been optically identified in NGC~6752
\citep{pool02b}, and 12-15 CVs have been identified in 47 Tuc
\citep{grin01,edmo03a,edmo03b}. \cite{bass04} also detect 2-3 CVs in the 
globular cluster M~4 in a similar manner.

\subsection{The formation of cataclysmic variables in globular clusters}
\label{sec:formcvs}

The formation mechanisms of CVs in globular clusters is currently
unclear.  In the field, collisions are rare and therefore CVs evolve
chiefly from their primordial binaries, where the more massive of the
two stars evolves more quickly.  It can then fill its Roche lobe and
unstable mass transfer can occur which can lead to a common envelope
system in which the stellar remnant and its companion lose their
angular momentum and can form a contact binary \citep{pacz76}.
However, in a globular cluster, such a primordial binary may have a
very wide separation and during an interaction with another cluster
star and/or binary, the binary can be broken up, especially if it is
located in a dense region, such as the core of the cluster.
Alternatively the initial binary may be harder, but eventually harden
to such an extent, through successive encounters, that the ultimate
fate of the system may be changed.  Instead the binary could undergo
an exchange encounter
\citep{davi97}.  Alternatively, a single white dwarf remaining 
after a star dying in a planetary nebula can encounter a cluster star.
This encounter can raise tides on the cluster star and transfer energy
and angular momentum into the envelope.  If sufficient energy is
transferred, the two objects can either merge or form a binary system,
a potential CV \citep{fabi75}.

Mass segregation concentrates binaries to the
centre of cluster, although the less massive binaries will not
necessarily be located in the core of the cluster
\citep{davi97}.  Indeed the distribution of the binaries depends
essentially on the central density of the cluster.  A higher core
density will yield a larger number of encounters, increasing the
number of CVs formed through encounters and at the same time decreasing
the primordial CV population.  Primordial CVs should therefore be
found preferentially outside the core and encounter CVs within
it.

Unlike the qNSLMXBs that have limiting luminosities of
$\sim$10$^{32}$ ergs s$^{-1}$, CVs can have luminosities of
$\sim$10$^{29}$ ergs s$^{-1}$ (0.5-2.5 keV) or even fainter
\citep{verb97}.   Therefore, unlike the population of qNSLMXBs 
where we have (almost) a complete population for each globular cluster
that has been observed, we do not have a complete population for the
CVs. Indeed \cite{dist94} find from cluster simulations of 47~Tuc and
$\omega$ Centauri, that there should be more than a hundred
CVs in each of these two clusters alone.  Without
X-ray spectra and in many cases follow-up optical photometry and
spectroscopy, we can not differentiate the CVs from other faint X-ray
sources in the globular clusters, such as active binaries.
\cite{hein03} noted that the majority of fairly hard sources with
luminosities greater than 10$^{31}$ ergs s$^{-1}$, found in the
centres of globular clusters by {\em Chandra} have been revealed
through optical photometry and spectroscopy to be CVs.  They then
undertook a similar analysis on this hard population (after removing
sources that were already known not to be CVs) as to that undertaken
for the qNSLMXBs.  They found that for this population, the results
were not in agreement with the theoretically predicted formation rate
from close encounters and that there was a weaker dependence on the
central density than in the formation of qNSLMXBs.  They propose that
high-density environments preferentially destroy CVs.  However, they
note exceptions to this assertion.  NGC~6397, the densest globular
cluster studied in this analysis, has an apparent excess of X-ray
sources \citep{pool03}, which suggests the opposite
conclusion. However, \cite{hein03} state that NGC 6397 may be an
unusual cluster, being possibly disrupted.  They also state that the
formation of some CVs in globular clusters from primordial,
undisturbed binaries may partially explain the weaker dependence on
density. However, there are many different ways that a globular
cluster can be disturbed, which can change the formation/evolution of
the cluster members \citep[see e.g.][]{webb04b,gned97}.

\subsection{Where are the cataclysmic variable outbursts?}

One of the most striking differences between globular cluster CVs and
field CVs is the lack of outbursts observed in globular cluster CVs.
Many types of field CVs (U~Gem, SU~UMa, Z~Cam etc) show outbursts
every few weeks to months.  However, the evidence for globular cluster
CV outbursts is poor.  Two eruptions of the CV V2 in the globular
cluster 47~Tuc have been reported \citep{pare94,shar96}.  Through
imaging of the globular cluster M~5, \cite{shar87} observed what they
proposed was an outburst and decline to quiescence of a CV identified
as V101 \citep{marg81}.  \cite{nayl89} confirmed the CV posit through
optical observations in outburst and quiescence.  Further, searches for
evidence of the existence of CVs in globular clusters have often had
little success. \cite{ciar90} searched for nova outbursts, as
indicators for the presence of CVs in 54 of M~31's globular
clusters. Over an effective survey time of about 2.0 years, no cluster
exhibited evidence for a nova explosion.  Moreover, indications for CV
outbursts have often been repudiated.

It is unclear why there should be so few CV outbursts observed, as
there is no reason to suggest that CV activity should be suppressed in
globular clusters.  \cite{shar96} suggests that tidal capture may lead
to runaway mass transfer in almost all cases, which would explain the
lack of CVs observed.  Alternatively low accretion rates may result
and the time between outburst could be long.  Following the detection
of two CVs in an {\em open} cluster, where one was in outburst,
\cite{kalu97} noted that it would be difficult to identify a CV
undergoing a similar type of outburst in a globular cluster using the
two methods most often applied to look for CVs in globular cluster
centres (searching for photometric variability or selecting objects
with strong Balmer emission).  This may explain why few CVs in
outburst have been detected using optical observations.

It has been suggested that the CVs detected by X-ray observatories in
globular clusters may be primarily magnetic CVs c.f. the 5 magnetic
CVs discussed in \cite{grin99} that were detected initially by {\em
Rosat}.  Magnetic CVs have high L$_x$/L$_{opt}$ values, do not show
outbursts because their accretion discs are disrupted by the magnetic
fields and show X-ray spectra which are particularly hard due to
magnetically channeled accretion, which makes them more easily
detected in X-rays \citep{patt94}.  Indeed, the CVs observed by {\em
XMM-Newton} for which we have X-ray spectra all have best fits that
are consistent with very high temperature bremsstrahlung models,
expected of magnetic CVs.  This could explain why the CVs detected in
X-rays have not been seen to outburst.

\section{Millisecond pulsars}
\label{sec:msps}

It is thought that millisecond pulsars (MSPs) are the descendants of
NSLMXBs.  During the accretion phase, it is believed that there is a
transfer of angular momentum onto the neutron star, spinning it up to
high velocities.  Particles close to the surface of the neutron star
are accelerated out along the open field lines, so we observe a beam
of radiation as the neutron star rotates, most readily observable in
the radio domain, but also observable at other wavelengths such as in
X-rays \citep{prin72}.  Many millisecond pulsars found in globular
clusters prior to 1999 when the observatories {\it XMM-Newton} and
{\it Chandra} were launched, were detected by radio telescopes
\citep[e.g.][]{cami00,dami01}.  However, both X-ray observatories and
notably {\it Chandra} have been efficient in detecting a large
population of globular cluster MSPs.

Millisecond pulsars can also have soft X-ray spectra like the
qNSLMXBs.  These spectra are well fitted by low temperature
blackbodies or hydrogen atmosphere models, where the temperature and
radius of the emission are those of the heated polar cap
\citep[10$^6$-10$^7$ K and $\sim$1 km e.g.][and references
therein]{zhan03,zavl98}.  MSPs can also have spectra
that are well fitted by hard power laws, where the emission is due to
particles accelerated in the magnetosphere or spectra composed of the
two different types of emission \citep[e.g.][]{webb04c,webb04d}.
These objects have lower luminosities than qNSLMXBs, unless
there is accretion still taking place in the system.

MSP populations in globular clusters have been shown to be extremely
diverse.  \cite{grin02} state that the implied population of MSPs in
47~Tuc, using {\it Chandra} observations, is between 35 and 90.
However they note that NGC~6397 is relatively deficient in MSPs, with
only one MSP detected using radio and/or X-ray observations.  They
also discuss the fact that this MSP lies at a radius of 11 core radii
from the centre, which indicates that it may have been ejected from
the core, perhaps in a second exchange encounter.  Many of the MSPs
observed with {\it Chandra} have been detected with only a few counts,
not enough to extract a spectrum.  An X-ray colour analysis has
indicated that the (majority of the) sources have very soft X-ray
spectra, exploited to identify their MSP nature.  Again, an advantage
of the {\it XMM-Newton} data is that we can obtain spectra of the
MSPs, which can help to distinguish MSPs showing hard spectral
emission from other sources showing hard spectra e.g. CVs \citep[two
MSPs in the globular cluster M~55][]{webb04b}.

\cite{grin02} showed that one of the key results in their study was 
that MSPs in the globular clusters 47~Tuc and NGC~6397 may have a less
efficient conversion of rotational spin-down energy into soft X-rays
than most field MSPs \citep[see][]{beck97}.  Why this should be is
unclear.  \cite{grin02} suggest that these pulsars may have an altered
surface magnetic field (and thus an altered light cone radius) or
possibly larger masses (or compactness).  They propose that MSPs in
dense cluster cores, unlike those in the field, have the possibility to
be driven back into contact and thus into an accretion phase, as a
recycled NSLMXB.  Renewed accretion could continue to bury the
magnetic field, a process thought to be responsible for field decay
observed from neutron stars at birth to the values typical of MSPs
\citep{roma90}. This could lead, in turn, to an altered magnetic field
configuration, particularly for a secondary re-exchanged into an MSP
(with random spin-orbital encounter angular momentum). This method
would also systematically increase the neutron star mass beyond that
for field MSPs.  There is, however, some opposition to the field
burial hypothesis e.g. \cite{rude91} who proposes that during spin up,
the surface magnetic field evolution mirrors the changes in the core
magnetic field configuration.  The surface magnetic field can migrate
to the pole of the hemisphere it started in, which can account for
both the change in the magnetic field configuration and its strength.
Further, \cite{chen03} suggest that a small-scale, strong, surface
magnetic field may play a role in determining the X-ray emission
properties of MSPs, where the existence of such a magnetic field may
depend on the formation history, possibly providing an explanation for
the differences between the MSPs in the field and in globular
clusters.  It is unclear which hypothesis is correct, but the study of
globular cluster MSPs should help resolve the physical nature of their
magnetic fields.

Another pertinent question pertains to the populations of MSPs in
globular clusters.  If MSPs are born from NSLMXBs, whose numbers
appear to be correlated with the encounter rate of the cluster, where
are all the MSPs in clusters with high encounter rates, such as
NGC~6440 \citep{pool02} which has shown evidence for very few MSPs?
\cite{grin02} propose that some of the main-sequence or BY Dra
binaries may harbour MSPs, such as in the case of
the MSP 6397A in NGC~6397.  Alternatively, this could be related to
the problem of the retention of neutron stars in globular clusters,
see e.g. \cite{pfah02} or simply that the observations are
insufficient to detect all the MSPs.

\section{Outlook for the future}

Both {\it XMM-Newton} and {\it Chandra} are expected to remain in
service for several years to come.  As outlined here, they have
already revolutionised our understanding of faint X-ray sources in
globular clusters.  Not only have we confirmed the nature of these
sources, we have identified populations of binaries and their
descendants, that have already and will continue to assist us in
understanding their formation and evolution.  Future observations with
these observatories, coupled with multi-wavelength follow-up
observations, should not only increase the population sizes, but also
enhance our understanding of these populations.

As outlined in Section~\ref{sec:nslmxbs}, we expect to eventually
place strict constraints on the neutron star equation of state through
studying the NSLMXBs in globular clusters, as well as understand
accretion processes in the quiescent systems.  We also expect to be
able to understand the nature and origin of the hard power law
component observed in the X-ray spectra of some of the qNSLMXBs.
Studying the optical counterparts to these systems, which are now
beginning to be discovered, will also enhance our understanding of
their origin.

Further observations should eventually provide observational support
for the formation mechanisms of CVs in globular clusters, as well as
understand the possible differences between globular cluster and
field CVs.

With regards to MSPs, further observations should resolve the
diversity puzzles: Why do globular cluster MSPs have soft X-ray
spectra indicative of heated polar caps, but many field MSPs show hard
X-ray spectra, indicative of particles accelerated in the
magnetosphere?  Why do globular cluster MSPs have a less efficient
conversion of spin-down energy into X-rays than most field MSPs?
And what can explain the varying globular cluster populations of MSPs?

\end{document}